\documentclass[preprint2]{aastex}
 
\usepackage{amssymb}     
\usepackage{epsfig}
 
\newfont{\sfsl}{cmssqi8 scaled 1200}
\newfont{\sfslp}{cmssqi8 scaled 1250}
\newfont{\sfsls}{cmssqi8 scaled 900}
\newfont{\sfsln}{cmssqi8 scaled 1350}
\newfont{\sfslz}{cmssqi8 scaled 1500}
\newfont{\sfslzz}{cmssqi8 scaled 1150}
\newfont{\sfslms}{cmssqi8 scaled 1000}	
\newfont{\sfsla}{cmssqi8 scaled 3000}
\newfont{\sfslb}{cmssqi8 scaled 2200}
\newcommand{\gcsa}{{\sfsl HIFLUGCS}} 

\newcommand{\ra}{{\rm RASS}}
\newcommand{\ps}{{\rm PSPC}}
\newcommand{\as}{{\it ASCA}}
\newcommand{\ei}{{\it Einstein}}
\newcommand{\xs}{{\rm XSPEC}}
\newcommand{\xmm}{{\it XMM-Newton}}
\newcommand{\xmmbf}{{\bf XMM-Newton}}
\newcommand{\cha}{{\it Chandra}}

\newcommand{\om}{\Omega_{\rm m}}

\newcommand{\fx}{f_{\rm X}}

\newcommand{\cx}{C_{\rm X}}

\newcommand{\nh}{N_{\rm H}}

\newcommand{\tx}{T_{\rm X}}

\newcommand{\eb}{0.1-2.4\,{\rm keV}}

\newcommand{\esc}{\times10^{-11}\,{\rm ergs\,s^{-1}\,cm^{-2}}}

\shorttitle{\xmm\ Observation of A1644}

\begin{document} 
 
\title{\xmmbf\ Observation of the Merging Galaxy Cluster A1644} 

\author{\anchor{http://www.reiprich.net}{Thomas H. Reiprich} and Craig L. Sarazin}
\affil{\anchor{http://www.astro.virginia.edu}{Department of Astronomy,
University of Virginia}}
\affil{P.O. Box 3818, Charlottesville, VA 22903-0818, USA}
\email{thomas@reiprich.net, sarazin@virginia.edu}
\and
\author{Joshua C. Kempner}
\affil{Harvard-Smithsonian Center for Astrophysics}
\affil{60 Garden Street, MS-67, Cambridge, MA 02138, USA}
\email{jkempner@cfa.harvard.edu}
\and
\author{Eric Tittley}
\affil{Institute for Astronomy, University of Edinburgh}
\affil{Edinburgh, Midlothian, EH9 3HJ, UK}
\email{ert@roe.ac.uk}

\begin{abstract}
An \xmm\ imaging spectroscopy analysis of the galaxy cluster A1644 is presented.
A1644 is a complex merging system consisting of a main and a sub
cluster. A trail of cool, metal-rich gas has been discovered close to the sub
cluster. The combination of results from X-ray, optical, and radio data, and a
comparison to a hydrodynamical simulation suggest that the sub cluster has
passed by the main cluster off-axis and a fraction of its gas has been stripped
off during this process.
Furthermore, for this merging system, simple effects are illustrated which can
affect the use of clusters as cosmological probes.
Specifically, double clusters may affect estimates of the
cluster number density when treated as a single system. Mergers, as well as cool
cores, can alter the X-ray 
luminosity and temperature measured for clusters, causing these values to differ
from those expected in equilibrium.
\end{abstract}


\section{Introduction}\label{intro}

Clusters of galaxies are formed in violent events, cluster mergers.
These mergers can dramatically affect the observed properties of
clusters, particularly in the X-ray band
\citep[e.g.,][]{mgd02}.
Signatures of mergers include ``cold fronts''
(\citealt{mpn00}; \citealt*{vmm01b}),
merger shocks
\citep[e.g.,][]{mgd02},
and multiple cool cores
\citep*{msv99}.
The application of simple hydrodynamical analysis to these merger
signatures allows the merger Mach number and other aspects of the merger
kinematics to be determined (\citealt{vmm01b}; \citealt*{ksr02,s02}).
In general, merger Mach numbers are modest,
${\cal M} \sim 2$.

Mergers also provide interesting tests of the role of various physics
processes in clusters.
For example, very strong limits have been placed on the rate of thermal
conduction and diffusion
(\citealt{etf00}; \citealt*{vmm01a}; \citealt{mmv03})
and
Kelvin-Helmholtz instabilities and viscosity
\citep{vmm01a}.
They also provide limits on the relative roles of thermal versus nonthermal
pressure sources in the intracluster medium
\citep{msv99}.
The role of merger shocks in accelerating relativistic particles can also
be studied
\citep{mv01}.
Recently, the relative distributions of gas, galaxies, and dark matter in
merging clusters have been used to limit the self-collisional cross-section 
of the dark matter
\citep{mcg03};
collisional dark matter has been suggested to explain the mass profiles in
dwarf galaxies and to resolve other problems
\citep[e.g.,][]{ss00}.

On the other hand, merger effects can complicate the use of clusters of
galaxies as cosmological probes.
Clusters provide a number of important cosmological tests,
mainly because they are
the only objects which are
large enough to represent a fair sample of
materials in the Universe, yet small enough to be relaxed
\citep[e.g.,][]{bah00}.
Most analyses of clusters for cosmology assume that they are equilibrium
systems;
for example, the masses of clusters are often determined from hydrostatic
equilibrium, or more simply by assuming an equilibrium relation between the
cluster X-ray temperature or luminosity and the cluster mass.
During major mergers, clusters are not in hydrostatic equilibrium.
Also, compression and heating during a merger can produce a large transient
increase in the
X-ray temperature and luminosity
\citep{rs01}.
If equilibrium relations are applied to merging clusters, their masses can
be greatly over-estimated
\citep*{rsr02}.
This can be particularly important for high redshift clusters;  
in flux limited samples, merging clusters can be over represented due
to their boosted X-ray luminosities.
It is important to understand these effects by studying relatively nearby
clusters where more detailed dynamical information can be determined.

\begin{figure}[thbp]
\psfig{file=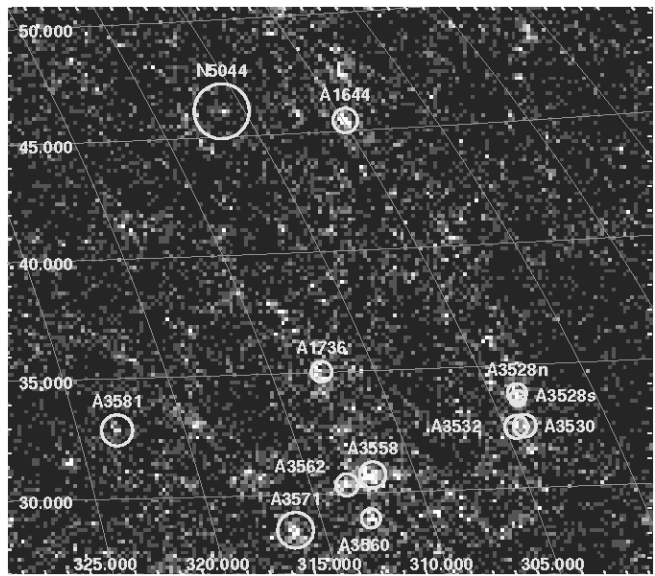,width=8.61cm,angle=0,clip=}
\caption{Location of A1644 relative to the core of the Shapley supercluster in
Galactic coordinates.
Shown are virial radii of a few clusters from \citet{rb01} on top of
a distribution of 2MASS galaxies.}\label{i2mass}
\end{figure}

In this paper, we present new \xmm\ observations of the Abell 1644 cluster.
Our analysis of the
\ei\ observation suggested that this cluster was undergoing a major merger.
\citet{jf99} classified A1644 as a double
cluster with roughly equal components based on the \ei\ image.
 From a deprojection analysis of the \ei\ data,
\citet*{wjf97} found that A1644 harbors a small cooling flow of about
12 solar masses per year.
\citet{gef97} noted the possible presence of two components in the
galaxy velocity distribution in the cluster, and listed two subclusters as
part of this cluster.
They also found that the cD galaxy had a large peculiar velocity, which
suggests the possible presence of minor substructure.
However, in the end they classified Abell 1644 as a regular cluster.
No evidence for substructure has been found in recent optical and near-infrared
observations by \citet{tgk01}.
A1644 is located close to a region of high galaxy cluster density. It is about 
15\,degrees north (in Galactic coordinates) of the core of the Shapley
supercluster (Fig.~\ref{i2mass}) at about
the same redshift, $z({\rm A1644})=0.0474$
\citep{1993AJ....106.1273Z}.

\section{Data Reduction and Analysis}\label{data}

\xmm, with its combination of high throughput, broad spectral band, good
spectral resolution, and fairly large field of view is ideal for studying
this merging system.
A1644 was observed by \xmm\ for a total duration of 23\,ks during revolution 199
(observation ID: 0010420201) on January 8, 2001.
The EPIC-MOS cameras were operated in Full Frame (FF) mode and the EPIC-pn
camera in Extended Full Frame (EFF) mode. The Thin1 filter was used in all EPIC
observations for protection against optical light contamination. Calibrated
event files have been created with \anchor{http://xmm.vilspa.esa.es/sas/} {SAS
V5.3}. The calibration index file has been created on May 7, 2002 based on the
most up to date calibration files available at that time. Most further reduction
and analysis steps have been performed using SAS V5.3 and the package
\anchor{http://heasarc.gsfc.nasa.gov/docs/software/lheasoft/} {HEAsoft V5.2}.

\subsection{Flare Rejection}\label{flare}

\xmm\ observations in general suffer from periods of very high background similar to
\cha\ observations \citep[e.g.,][]{lwp02}.
There are several possible ways to `clean' a data set of these flares.
Our method is based on the fact that we will use independent blank-sky
observations
(\S~\ref{backg})
to determine the main non-flare hard background component, the `Particle
Induced Background' (PIB).
(The PIB is distinct from the flares, which are believed to be produced by
`soft' protons.)
Thus, we must be able to clean the blank-sky background observations in
the same way as the source observation.
When determining the lightcurve this requires
ensuring a negligible `contamination' by emission from
astrophysical sources because otherwise the sensitivity to flare detection
varies between source and background observation.
Here we choose for the MOS cameras the energy
band 10--12\,keV and for pn the band 12--14\,keV similar to previous works
\citep*[e.g.,][]{mnr02}, which are dominated by PIB (background
induced directly or through fluorescent instrumental
lines by high energy particles) and soft protons (if present). Note that this
approach has the drawback of low statistics and furthermore assumes that
the flare characteristics in this energy band are representative of those
at lower energies as well.
In \S~\ref{backg} we test the effect of increased statistics and the latter
assumption.

Times were binned in 100\,s intervals, which appears to
be a reasonable compromise between good temporal resolution and low noise level.
This time interval also is large compared to the read out cycles of the cameras.
Further only events with pattern $\le$ 12 and flag = 0 have been used to ensure
a high rejection of potentially distorting events such as caused by hot/flickering
pixels/columns or features close to CCD edges especially at the read out nodes.

\begin{figure}[thbp]
\psfig{file=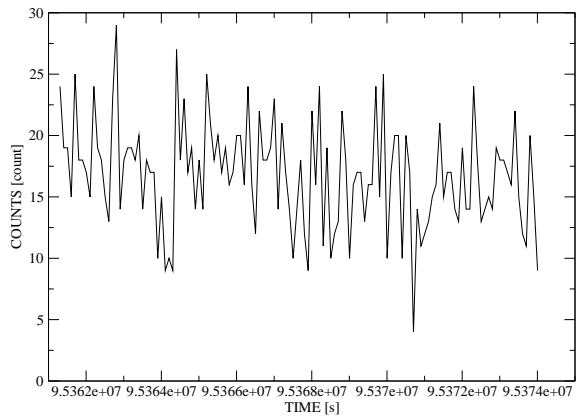,width=7cm,angle=270,clip=}
\caption{Lightcurve for pn with 100\,s binning.}\label{lcpn}
\end{figure}

\begin{figure}[thbp]
\psfig{file=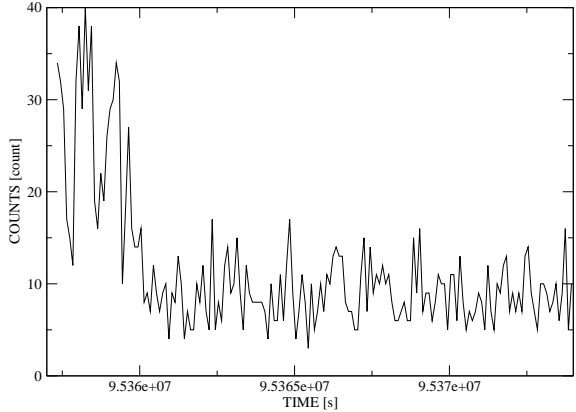,width=7cm,angle=270,clip=}
\caption{Lightcurve for MOS1 with 100\,s binning.}\label{lcMOS}
\end{figure}

Inspection of the resulting lightcurves showed no obvious flares in the pn data
(Fig.~\ref{lcpn}).
MOS count rates were higher by up to a factor $\sim$ 4 in the beginning
of the MOS observations compared to the remaining times (Fig.~\ref{lcMOS}). MOS
exposures started earlier than 
those for the pn camera; the lack of obvious flares in pn is consistent with
the MOS lightcurves for the overlapping time interval.
The obviously contaminated times in the MOS observations were
removed by taking into account only events with arrival times $\ge$ 95360200\,s
(\xmm\ time) for all further steps.

Before calculating mean count rates in order to test for background flares,
conservative `generic' cuts have been applied on the count rates, $\cx$. For pn (MOS)
only times with $2\ (1) \le \cx \le 45\ (25)$ counts per 100\,s have been
used. The lower threshold is especially important for pn data in order not to
bias the calculation of the mean count rate by inclusion of counting mode times,
times during which transmission of information for individual events of one or
more (most likely all four) quadrants to the ground station does not occur.

For the final cuts,
upper \emph{and} lower thresholds are required in order to treat
statistical fluctuations in both directions symmetrically. Here we choose to
define the good time intervals (GTIs) by including only times where the
count rates lie within $\pm$ 2\,$\sigma$ of the mean count rates.
For the individual detectors the count rate limits for GTIs are found to be
8--25 (pn), 3--15 (MOS1), and 2--15 (MOS2) counts per 100\,s.
The effective exposure time (live time) after this procedure results in 11\,ks
for pn and 13\,ks for MOS1 and MOS2.

\subsection{Vignetting Correction}\label{vigne}

Depending on the nature of a detected event one should take into account
vignetting caused by the baffles, mirrors, Reflection Grating Array (for
the MOS cameras only), etc.\ in the analysis. For instance the X-ray photons from the
target source (A1644) and the Cosmic X-ray Background (CXB) are vignetted in the
same way. The soft protons show a different vignetting characteristic (but
hopefully we got rid of most of those in \S~\ref{flare}) and the PIB is not
vignetted at all if the spatial variation
of the fluorescent lines caused by the camera material itself is not
interpreted as a vignetting characteristic. The noise in the CCDs, most easily
visible as low energy events close to the CCD edges, is not vignetted, either.

In general, however, it's not easily possible to determine the nature of an
individual event.
The
combination of methods for flare rejection, vignetting correction and background
subtraction employed here aims at a proper treatment of all components for the
spectral analysis. This
strategy involves correcting \emph{all} events for the vignetting expected for
X-ray photons regardless of their nature at this point.

The vignetting depends on detector position and energy. The vignetting
correction has been accomplished with the SAS task evigweight which adds a
weight column to the events file. The value of the weight is determined for each
event by taking into account the detector position and energy of the event.
Weights are assigned relative to the on-axis effective area so the weights have
values $\ge$ 1. To accumulate images and spectra these weights are summed up,
and errors are propagated including the vignetting correction.

\subsection{Background Subtraction}\label{backg}

In general, proper background subtraction is vital for an accurate spectral
analysis of extended sources.
In our case, the cluster emission fills most of the fields-of-view of
the detectors, so it is difficult to determine the background from
our observations.
Thus, we determine the background from so called blank-sky observations.
The blank-sky observations are
mostly high galactic latitude observations of regions with very few bright
sources. Several such observations have been combined for the EPIC cameras
mostly for the Thin filter in (mostly) FF mode \citep{lwp02} to achieve good
statistics and
reduce the effects of cosmic/sample variance.
We have made sure that for the energy range and signal-to-noise ratio in
spectral fits here, the differences in Charge Transfer Inefficiency (CTI)
corrections between the pn FF and EFF mode are not significant.
The blank-sky observations are available at the
\xmm\ calibration web
page\footnote{http://xmm.vilspa.esa.es/external/xmm\_sw\_cal/calib/epic\_files.shtml}.

The background observations have been recast on the sky in order to match
the source observations using the SAS task attcalc\footnote{See Andy Read's
scripts at http://www.sr.bham.ac.uk/xmm3/\,.}. This allows
using sky coordinates for regional selections and at the same time estimating
the background from the same detector regions as the source, which in turn ensures
that the significant variation of the fluorescent line emission with detector
position is properly taken into account.
By treating the blank-sky backgrounds identically to the data,
the effect of incorrectly applying a vignetting correction to the PIB
(in the source \emph{and} background
observations) is removed.

The same flare rejection method (\S~\ref{flare}) has been applied to the
background observations.
The count rate limits for GTIs here are found to be
8--31 (pn) and 2--16 (MOS1 and MOS2) counts per 100\,s. 

The PIB shows variation of about 8\% (1\,$\sigma$) as well as a long term
decrease of about 20\% between the beginning of \xmm\ observations and January
2001 \citep{kti02}. In order to account for these variations, total
field-of-view count rates of source and background observations have been
calculated using the same energy, pattern, and flag selection criteria as for
the GTI determination
(\S~\ref{flare}). The source to background count rate ratios $0.99\pm0.03$
(MOS1), $0.92\pm0.03$ (MOS2), and $0.88\pm0.02$ (pn) have been found
(1\,$\sigma$ errors), consistent with the expected
variation. For the calculation of these PIB scaling factors the events have not
been corrected for vignetting.

As mentioned in \S~\ref{flare}, cleaning of flares in the high energy bands generally used
does not necessarily imply that all flares have been removed.
Actually, we did find some weak remnant flares in the blank-sky observations
when plotting lightcurves (GTI screened as described in
\S~\ref{flare}) in broad energy bands \citep[see also][]{lwp02}. We therefore tested the variation in
(10--12\,keV) source to background count rate ratios when calculating GTIs in
the energy bands 7--12 and 0.3--12\,keV for MOS1 and MOS2. We found that the
variation is less than the (1\,$\sigma$) statistical uncertainty and
conclude that the influence of remnant flares is negligible for calculation of
background scaling factors here.

To check whether the limitation to high energy bands for the calculation of
background scaling factors affects the results significantly we have calculated
them also using the energy ranges 0.3--14\,keV (pn) and 0.3--12\,keV (MOS). In
this case, of course, only the CCD parts shielded from X-ray photons (and soft
protons) are usable.
We found results consistent with the total field-of-view results for pn and MOS1
and a slightly larger value for MOS2 
($1.00\pm0.02$).
In order to exclude the possibility that an imperfect correlation between the
normalization of the continuum and the fluorescence lines (dependent on detector
position) for the PIB caused this slightly larger than expected variation,
we also used the energy range 2--7\,keV to calculate the background scaling
factors. The results were consistent with the broader energy range used
previously.

In summary,
following these tests the background scaling factors 0.96 (MOS) and 
0.88 (pn) have been assigned to the observations. 
Most of the tests above yield results consistent with the expected statistical
uncertainties.
To account for the difference between MOS and pn
scaling factors
a conservative 8\% systematic error has been added to the full energy range
of all background spectra.
Note also that it is implicitly assumed here (and in most other works
that follow similar procedures) that the spectrum of the PIB does
not vary when its normalization changes.
Currently, there
seems to be no strong evidence for significant spectral variability
\citep[e.g.,][]{kti02} apart from the fluorescence lines.

The main drawback of this method is that in general one gets the background
contribution from the CXB wrong. The PIB dominates the hard part of the total
background but the CXB is important for the lower energies. Ideally one would
estimate the difference in CXB contribution to source and background
observations from a region with negligible source emission in the source
observation or another observation very close by
(e.g., \citealt*{mv01,paa01}; \citealt{mnr02,aml02}). This is not feasible in the case of
A1644 since cluster emission fills the entire field of view and the only other
\xmm\ observation within 5 degrees to date is likewise filled entirely with
NGC~5044's emission. Here, we model a possible CXB difference in the
spectral fitting procedure.

In order to estimate the spectral shape of a possible CXB difference between the
source and background spectra we have extracted spectra
from an irregular region with decreased cluster emission in the north-western
part of the observation and fitted a two temperature  
model. A high temperature component which is supposed to account for
the cluster emission and a low temperature, zero redshift, solar abundance
component which should represent the CXB excess relative to the background
observations.
The model fits indicate that in this region the cluster emission still
dominates. The ratio of the number of cluster photons to CXB photons equals
about 4 in the energy range 0.3--10\,keV and about 2 in the energy range
0.3--0.5\,keV.  
A best fit value $\sim 0.2$\,keV is found for the low temperature component.
This estimate is consistent with values from studies of soft
excess emission in other clusters \citep[e.g.,][]{klt03}.
The origin of this component is uncertain;
\citet{klt03} argue that it is due to intergalactic gas located around
clusters.
Our aim here is not to try to discuss
the source of this soft CXB excess, but rather to determine
its contribution to the cluster emission and account for it in the
model fits (\S~\ref{spect}).

\subsection{Spectral Fitting}\label{spect}

Spectra from MOS1, MOS2, pn have been fit simultaneously. The same pattern and
flag selections have been applied as for the generation of GTIs, except that
only single and double events have been used for the pn camera.
The response matrices were
m11\_r7\_im\_all\_2000-11-09.rmf, m21\_r7\_im\_all\_2000-11-09.rmf, and
epn\_ef2\\ 0\_sdY9.rmf for MOS1, MOS2, and pn respectively,
in addition to on-axis ancillary response files created with the SAS task
arfgen.

A comparison of individual pn single and
double event spectral fits gives consistent results.
The response of the pn camera depends on detector position.
However, it has been verified that for the
accuracy needed here, use of the epn\_ef20\_sdY0.rmf response matrix
does not result in significant changes of best fit parameter values.
Individual fit results from the three cameras are consistent with each other.
Point sources and hot pixels have been excluded.

\xmm's point spread function (PSF) is much improved compared to the PSFs of
previous satellites capable of spatially resolved spectroscopy, e.g., \as.
Nevertheless, for objects with strong temperature and surface brightness
gradients its effect can become important. However, the main emphasis in this
work is on the comparison of different segments of spectra at the same
radii from cluster centers. In this case radial PSF effects should not cause
any artificial differences.

Spectral fits have been performed within \xs\ using a model
(wabs*mekal)+mekal.
Parameters for the second thermal component have been fixed at a redshift
of zero,
a metal abundance of $A=1$ solar \citep{ag89}, and temperature of
$k\tx=0.2$\,keV, as discussed in \S~\ref{backg}.
The normalization is determined from the fits.

Source spectra have been regrouped to have at least 50 counts per bin. Spectral
fits have been performed nominally in the energy range 0.3--10.0\,keV 
for MOS and 0.3--12.0\,keV for pn. All errors are quoted at the 90\%
confidence level for one interesting parameter unless noted otherwise.

\section{Results}\label{resul}

\subsection{Imaging}\label{imagi}

\begin{figure}[thbp]
\psfig{file=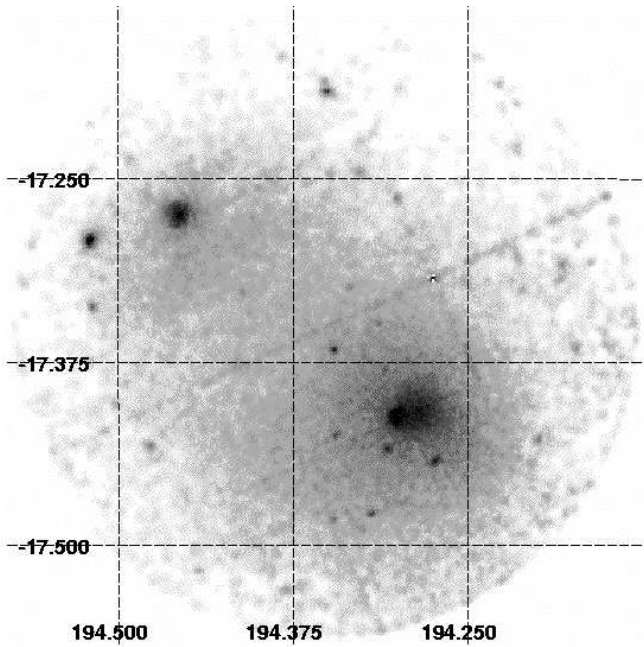,width=8.6cm,angle=0,clip=}
\caption{Background, exposure, and vignetting corrected,
adaptively smoothed, combined MOS1-MOS2-pn count rate image of A1644 in the
energy range 0.3--5.0\,keV (R.A. and Dec.\ in equatorial coordinates, epoch
J2000.0). 
The linear artifacts are due to an
imperfect exposure correction near the chip boundaries.}\label{ixmm}
\end{figure}

Figure~\ref{ixmm} shows the combined MOS1-MOS2-pn image of nearly the
full \xmm\ field-of-view of A1644.
The cluster has
a very complicated surface brightness distribution on all
scales.
A main cluster to the southwest and a smaller sub cluster to the northeast are
easily identified.
The emission surrounding both core regions is highly non-spherical.
There is stronger emission to the south of the sub clump than to the north.
The core region of the main clump itself contains a displaced
core-within-a-core. 
\begin{figure}[thbp]
\psfig{file=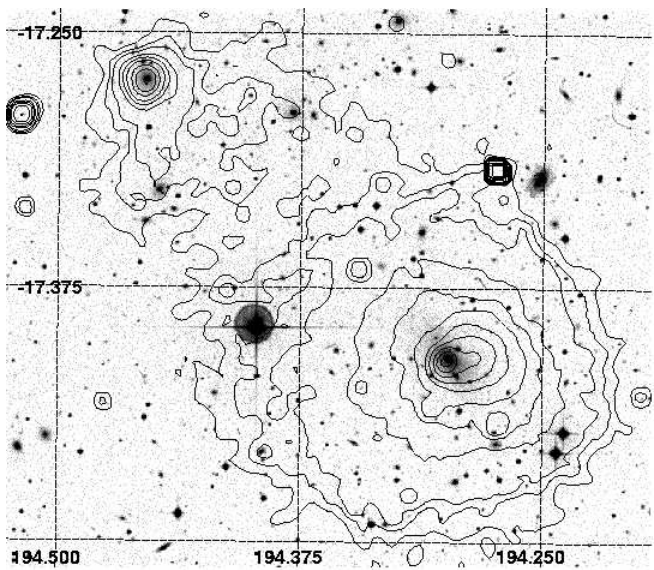,width=8.6cm,angle=0,clip=}
\caption{X-ray contours from the image in Fig.~\ref{ixmm} overlaid onto an
optical DSS image.}\label{iopt}
\end{figure}

The Digitized Sky Survey (DSS) image in Fig.~\ref{iopt} shows that the X-ray
peaks of main cluster and sub cluster coincide with two Bright Cluster Galaxies
(BCGs).
The relative line-of-sight velocity component of the two central
galaxies is about 600\,km\,s$^{-1}$.
The X-ray contours to the north of the sub clump appear to be slightly
compressed.
Radio observations of the two BCGs show that they are both
radio sources \citep[e.g.,][]{ol97}.
\begin{figure}[thbp]
\psfig{file=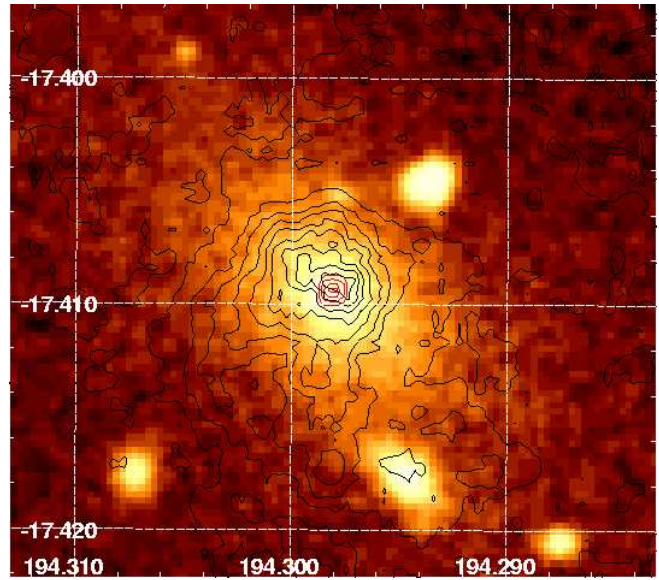,width=8.6cm,angle=0,clip=}
\caption{The DSS image of the inner region of the main clump, overlaid with
the X-ray contours from \cha\ (black) and, in the very center, the 1.4~GHz radio
data (red; from M. Ledlow).}\label{irad1}
\end{figure}
\begin{figure}[thbp]
\psfig{file=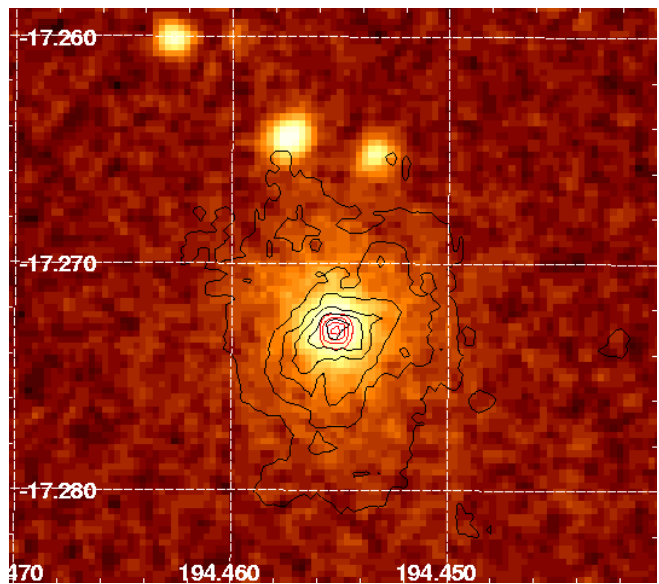,width=8.6cm,angle=0,clip=}
\caption{The DSS image of the inner region of the northeast sub clump
overlaid with
the X-ray contours from \cha\ (black) and, in the very center, the 1.4~GHz radio
data (red; from M. Ledlow).}\label{irad2}
\end{figure}

The regions around the two BCGs are shown in more detail in
Figs.~\ref{irad1} and \ref{irad2}. Included are X-ray contours of a 19\,ks \cha\
observation taken on the same day as the \xmm\ observation and available from
the public archive.  These contours illustrate the complexity of the emission in
the very cluster cores. Note that also in the very center of the sub clump the
X-ray contours to the north appear slightly compressed
(Fig.~\ref{irad2}).
This apparent compression is on a much smaller scale than the compression
already noted in Fig.~\ref{iopt}. 
Optical, radio, and X-ray centers coincide almost exactly for both sub clusters.

\subsection{Spectroscopy}\label{spectro}

The intracluster gas temperature is a fundamental observable for a
gravitational cluster mass determination based on the hydrostatic assumption.
The overall best fit temperature based on a large elliptical region centered at
R.A.\ $=194.356\deg$ and Dec.\ $=-17.363\deg$ encompassing
both sub clusters (semi major axis $=733\arcsec$, semi minor axis
$=528\arcsec$, position angle $=45\deg$) is $k\tx=3.83\pm0.06$\,keV. The best fit
metal abundance is $0.32^{+0.02}_{-0.03}$ solar for the same region.

Note that the column density of neutral Galactic hydrogen, $\nh$, has
been fixed at the value inferred from 21 cm radio measurements for our Galaxy
\citep[$5.33\times 10^{20}$\,cm$^{-2}$;][]{dl90} for all results.
In general this value is marginally larger but consistent with best fit values
from direct spectral fits. Best fit values for temperatures and abundances change
only well within their uncertainties if the column density is treated as a free
parameter.\footnote{Note that neglecting the CXB difference between source and
background observations, i.e., fitting a model wabs*mekal to the spectra here,
results in $\nh$ values significantly lower than the Galactic value, or, if
$\nh$ is fixed to the Galactic value, to artificially decreased temperatures and
increased $\chi^2$.}

\begin{figure}[thbp]
\psfig{file=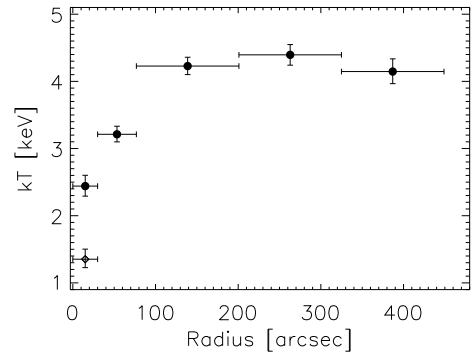,width=7.5cm,angle=0,clip=}
\caption{Radial temperature profile for the main clump. The open diamond shows the
best fit temperature if a third thermal component is included (see text).
}\label{Tprofm}
\end{figure}

\begin{figure}[thbp]
\psfig{file=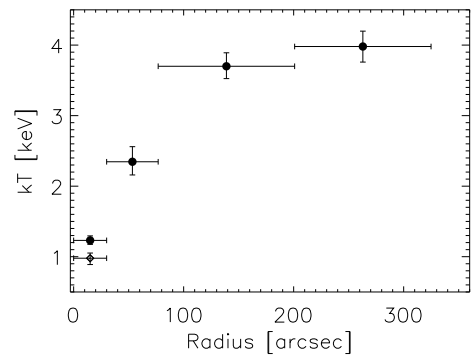,width=7.5cm,angle=0,clip=}
\caption{Radial temperature profile for the sub clump. The open diamond shows the
best fit temperature if a third thermal component is included.
}\label{Tprofs}
\end{figure}

Figures~\ref{Tprofm} and \ref{Tprofs} show the radial temperature profiles for
the main and sub clump, respectively.
Note that the center for the innermost bin
of the main clump is offset from the center of the other bins in order to
account for the core-within-a-core structure.
The reduced $\chi^2$ values (and the best fit $\nh$ values if left as a free
parameter) for the innermost bin of both sub clusters are too large
(Tab.~\ref{tbl:mprof} and \ref{tbl:sprof}). This is
likely caused by higher temperature gas in the line of sight. For these two
regions we therefore included a third thermal component, i.e., the model
(wabs*(mekal+mekal))+mekal was fitted. The limited statistics require freezing the
temperatures and abundances of the third components to the values found further
out. We therefore fixed the abundances at 0.5 solar and the temperatures at 3.5
and 3\,keV for the main and sub clump, respectively.
As a result the best fit central temperatures decrease further for both main and
sub clump (open diamonds).

Each of the two
temperature profiles appears surprisingly similar to temperature profiles of
relaxed, apparently undisturbed
clusters: a drop in the center to about 1/3 of the ambient gas
temperature, an isothermal structure in the outer parts, and weak
indications for a slight temperature drop in the very outermost regions
accessible.

Note that the addition of a third component is only a crude approximation to
account for effects caused by projection, PSF \citep[see, e.g.,][]{m02,mnr02},
and a possible real multiphase temperature structure. The
reduced $\chi^2$ values of the central bins become significantly smaller but
are still large even after addition of the third component. Also, these effects
need not be limited to the central bins alone. A1644 is an extremely irregular
object (see Figs.~\ref{ixmm} and \ref{HR}). This fact makes A1644 so interesting
but at the same time complicates deprojection attempts, for which
spherical symmetry usually is assumed.
In this work, we therefore focus on projected
temperature estimates and qualitative comparisons to temperature maps of
simulated clusters. 
\begin{figure}[thbp]
\psfig{file=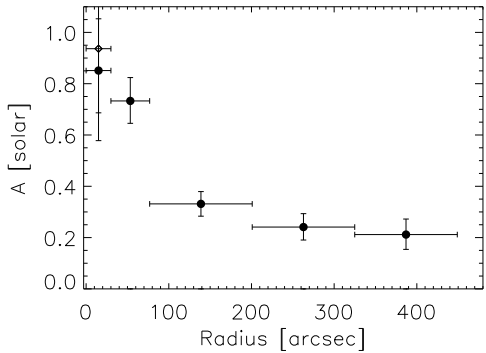,width=7.5cm,angle=0,clip=}
\caption{Metal abundance profile for the main clump. The open diamond shows the
best fit abundance if a third thermal component is included.
}\label{Aprofm}
\end{figure}
\begin{figure}[thbp]
\psfig{file=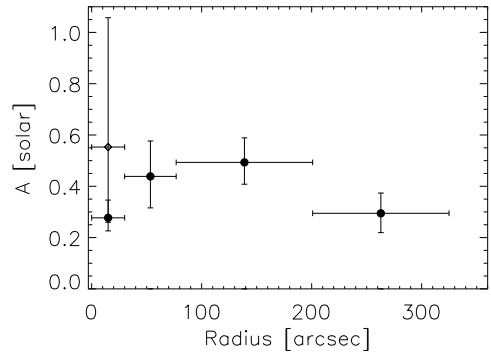,width=7.5cm,angle=0,clip=}
\caption{Metal abundance profiles for the sub clump. The open diamond shows the
best fit abundance if a third thermal component is included.
}\label{Aprofs}
\end{figure}

The corresponding metal abundance profiles are shown in Figs.~\ref{Aprofm} and
\ref{Aprofs}. 
In general, the abundances increase towards the centers of the two clumps,
although there seems to be a drop in abundance in the innermost region of the
sub clump.
However, this abundance drop vanishes when performing the fit with three thermal
components (open diamond). This drop can therefore be regarded as artificial,
even though it should be noted that the precise values depend on the
temperature and abundance values adopted for the third component.

It is worth noting that weak indications for higher normalizations of the CXB
excess emission towards the center of the main clump are found which may support
the findings of \citet{klt03}, although a discussion of the nature of the CXB
excess is beyond the scope of this paper (see \S~\ref{backg}).
%
\begin{deluxetable}{ccccc}
\tabletypesize{\footnotesize}
\tablecaption{Fit results for the main cluster profile \label{tbl:mprof}}
\tablewidth{0pt}
\tablehead{
\colhead{Radial range [\arcsec]} & \colhead{$k\tx$ [keV]}   &
\colhead{$A$}   & \colhead{red.\ $\chi^2$} & \colhead{D.O.F.}
}
\startdata
    0--30 & $ 2.44^{+ 0.16}_{- 0.15}$ & $0.85^{+0.20}_{-0.16}$ &  1.80 &  103 \\[0.9mm]
    0--30\tablenotemark{a} & $ 1.35^{+ 0.15}_{- 0.13}$ & $0.94^{+1.11}_{-0.36}$ &  1.46 &  103 \\[0.9mm]
   30--77 & $ 3.21^{+ 0.12}_{- 0.11}$ & $0.73^{+0.09}_{-0.09}$ &  1.19 &  312 \\[0.9mm]
   77--201 & $ 4.23^{+ 0.13}_{- 0.13}$ & $0.33^{+0.05}_{-0.05}$ &  1.11 &  769 \\[0.9mm]
  201--325 & $ 4.39^{+ 0.15}_{- 0.15}$ & $0.24^{+0.05}_{-0.05}$ &  1.09 &  808 \\[0.9mm]
  325--449 & $ 4.15^{+ 0.19}_{- 0.18}$ & $0.21^{+0.06}_{-0.06}$ &  0.97 &  872 \\[0.9mm]
\enddata
\tablenotetext{a}{Third thermal component applied (see text).}
\end{deluxetable}
%
\begin{deluxetable}{ccccc}
\tabletypesize{\footnotesize}
\tablecaption{Fit results for the sub cluster profile \label{tbl:sprof}}
\tablewidth{0pt}
\tablehead{
\colhead{Radial range [\arcsec]} & \colhead{$k\tx$ [keV]}   &
\colhead{$A$}   & \colhead{red.\ $\chi^2$} & \colhead{D.O.F.}
}
\startdata
    0--30 & $ 1.23^{+ 0.06}_{- 0.06}$ & $0.28^{+0.07}_{-0.05}$ &  2.65 &   54 \\[0.9mm]
    0--30\tablenotemark{a} & $ 0.98^{+ 0.07}_{- 0.09}$ & $0.55^{+0.50}_{-0.29}$ &  1.75 &   54 \\[0.9mm]
   30--77 & $ 2.35^{+ 0.21}_{- 0.19}$ & $0.44^{+0.14}_{-0.12}$ &  1.27 &  108 \\[0.9mm]
   77--201 & $ 3.70^{+ 0.19}_{- 0.17}$ & $0.49^{+0.10}_{-0.09}$ &  0.96 &  404 \\[0.9mm]
  201--325 & $ 3.98^{+ 0.22}_{- 0.22}$ & $0.29^{+0.08}_{-0.08}$ &  0.89 &  570 \\[0.9mm]
\enddata
\tablenotetext{a}{Third thermal component applied (see text).}
\end{deluxetable}
\begin{figure}[thbp]
\psfig{file=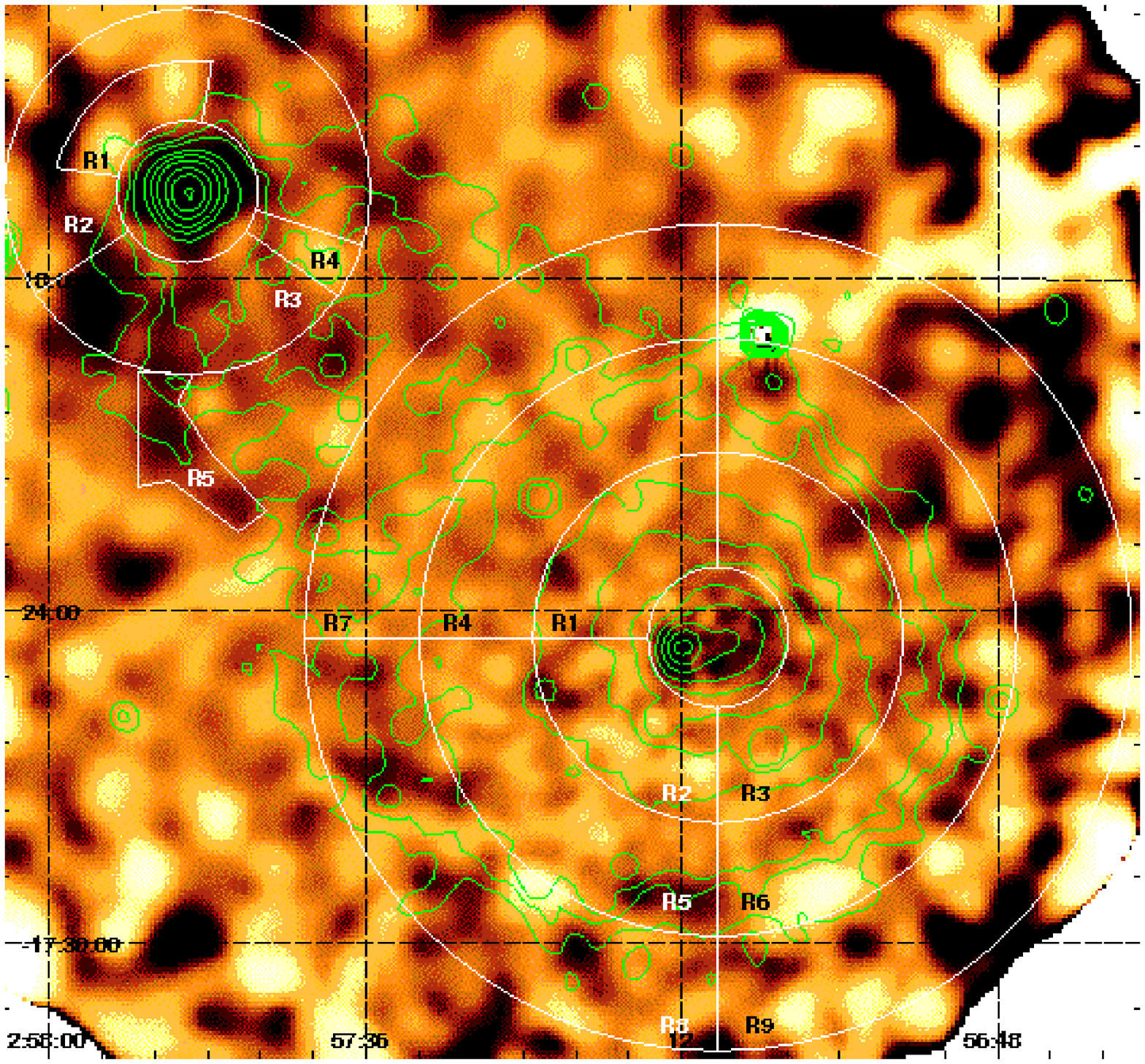,width=8.6cm,angle=0,clip=}
\caption{Hardness ratio map of A1644, based on the ratio of the
combined MOS1-MOS2-pn
count rates in the energy bands 0.3--2.0\,keV and 2.0--10.0\,keV.
The map has been corrected for background, exposure, and vignetting, and
is adaptively smoothed.
Soft emission appears dark and hard emission bright.
Also shown are surface brightness contours from Fig.~\ref{iopt} and regions
selected for spectral analysis.}\label{HR} 
\end{figure}

Figure~\ref{HR} shows a hardness ratio (HR) map. 
To create this map, first, source photon images from MOS1, MOS2, and pn (using
only single events) in the
energy band 2.0--10.0\,keV have been created and combined. Then, the
corresponding exposure maps have been created and combined, which also account
for vignetting.
Using the combined image and exposure map, an adaptively smoothed
and exposure and vignetting corrected source image has been created with the
SAS task asmooth (signal-to-noise ratio $=15$).
Subsequently, a similarly created
(though simply smoothed with $\sigma=7.5$), exposure and vignetting 
corrected, combined background image has been subtracted from the source image.
The same procedure has been followed to create an image in the 0.3--2.0\,keV
energy band, using the same smoothing templates as created for the 2.0--10.0\,keV
image. The ratio of the latter two images is shown in Fig.~\ref{HR}. Soft 
emission is shown in dark and hard emission in bright.
We have verified that using the source smoothing template also for smoothing the
background image does not result in significant changes in the relevant region.
Note that the CXB excess emission has not been subtracted in the HR map. This
may bias the HR map in the outermost regions of the observation. However, we
will find below that spectral fits that take the CXB excess into account
correspond well to the features seen in the HR map. Therefore, it is not
expected that the region shown is biased significantly.
\begin{figure}[thbp]
\psfig{file=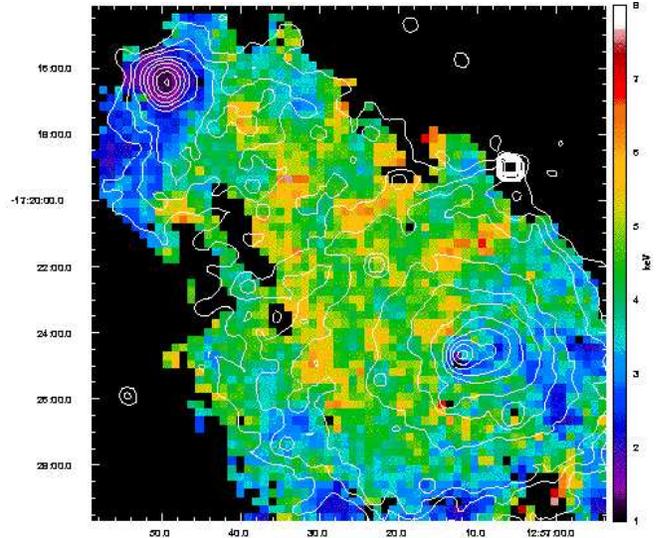,width=8.6cm,angle=0,clip=}
\caption{Temperature map of A1644.
Also shown are surface brightness contours from Fig.~\ref{iopt}.}\label{Txmap} 
\end{figure}

In Fig.~\ref{Txmap} an adaptively binned temperature map is shown.
The map has been created using all
three detectors and making
use of the Interactive Spectral Interpretation System \citep[ISIS; ][]{hd00}.
The pixel size is 14.4\arcsec . Pixels are nearly independent at the center of
the main clump but are correlated in other regions. The actual spectral
extraction region used for 
the fits depends on the surface brightness; the largest extraction region is    
1\arcmin on a side.
Most pixels shown black had fewer than 800 source counts using the maximum smoothing
scale. A few more pixels with negative error $>$ 30\% were masked out.
The abundance of the
main mekal component has been fixed here at the average value 0.32 solar
(\S~\ref{spectro}).
The features seen in the HR map (to be discussed in detail below) seem to
correspond to features in the temperature map. This indicates the
consistency of  
both approaches and we'll use the HR map, which is more closely related to the
data, as guidance below.

The hardness ratio map and the X-ray image suggest a number of interesting
regions in the cluster which may have differing temperatures and/or
abundances.
In general, 
clusters as irregular as A1644 often show more complex temperature
structure. Therefore, we also extracted spectra in segments of annuli for the
main clump.
The radial boundaries for the nine segments of the main cluster correspond to
the ranges for the three outermost annuli given in Tab.~\ref{tbl:mprof}. 
Figure~\ref{HR} shows these and other selected regions as well as
surface brightness contours overlaid onto the HR map.
(Note that the cluster emission extends beyond the outermost shown surface
brightness contour.)

\begin{figure}[thbp]
\psfig{file=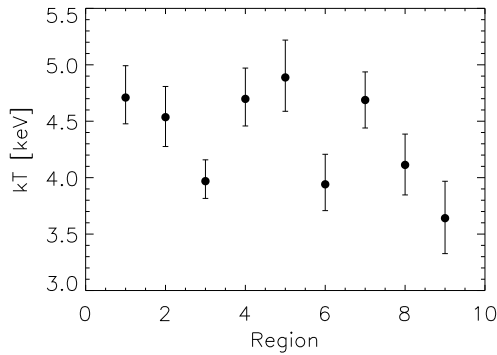,width=7.5cm,angle=0,clip=}
\caption{Gas temperature determinations of selected regions of the main clump
(see Fig.~\ref{HR}).}\label{Tregm}
\end{figure}

\begin{figure}[thbp]
\psfig{file=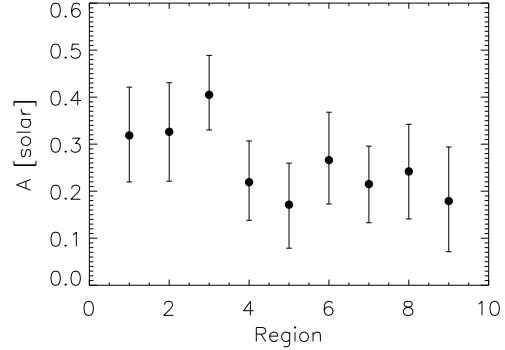,width=7.5cm,angle=0,clip=}
\caption{Metal abundance determinations of selected regions of the main clump.
}\label{Aregm}
\end{figure}

Best fit temperature and metal abundance values for
the segments are shown in Figs.~\ref{Tregm}
and \ref{Aregm} (and Tab.~\ref{tbl:mreg}), respectively. 
It is apparent that there is significant nonradial temperature structure,
and that the relatively regular radial temperature profile
in Figure~\ref{Tprofm} is the result of averaging over annuli with large
azimuthal temperature variations.
In particular,
the temperatures of the segments to the west of the main clump (R3,
R6, and R9) are all significantly lower than all regions to the east
(R1, R2, R4, R5, and R7) except for R8.
These findings correspond well to those in the HR map (Fig.~\ref{HR}). 
\begin{figure}[thbp]
\psfig{file=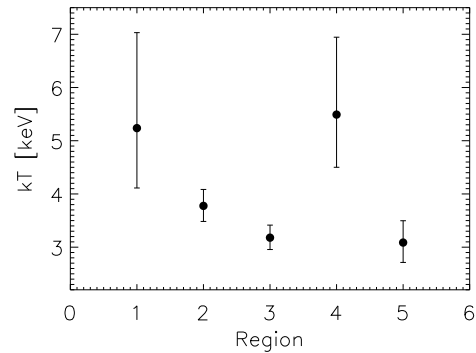,width=7.5cm,angle=0,clip=}
\caption{Gas temperature determinations of selected regions of the sub clump.
}\label{Tregs}
\end{figure}
\begin{figure}[thbp]
\psfig{file=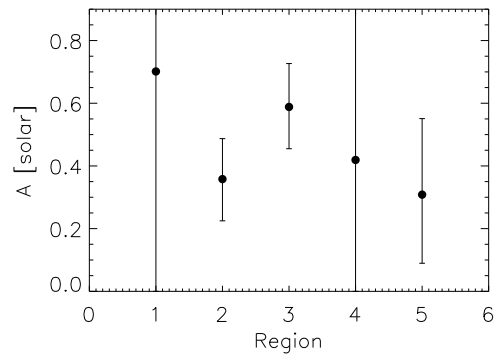,width=7.5cm,angle=0,clip=}
\caption{Metal abundance determinations of selected regions of the sub clump.
}\label{Aregs}
\end{figure}

We also determined temperatures and abundances for more complicated
regions selected as being of interest based on the \xmm\ image and the HR
map.
This allows us to search for cool/metal-rich trails (dark in the HR map)
or hot spots (bright in the HR map).
Whether brightness fluctuations in the HR map correspond to significant
temperature/abundance changes is evaluated by direct spectral fits of selected
regions. 
Note that not all artifacts, e.g.,
inexact exposure correction close to CCD chip boundaries, have been removed in
the HR map in Fig.~\ref{HR}.
The spectral analysis, however, is not affected by this.

The region to the south of the sub clump (R3) appears fairly soft,
whereas a small region to the southeast (R4) appears hard.
The spectral fit results (Fig.~\ref{Tregs} and Tab.~\ref{tbl:sreg}) reveal that
indeed the temperature determination for R3 is 
significantly lower than the value for the rest of this annulus (R2, which
excludes R1) and especially than that for R4.
Furthermore, the metal abundance of R3 appears enhanced (Fig.~\ref{Aregs}).
The best fit temperature for R5 (the small dark tail to the
south of R3 in Fig.~\ref{HR}) is similar to that for R3.
On the other hand, R1, the
region north-northeast of the sub clump, has a high temperature
similar to R4.
%
\begin{deluxetable}{ccccc}
\tabletypesize{\footnotesize}
\tablecaption{Fit results for the main cluster regions \label{tbl:mreg}}
\tablewidth{0pt}
\tablehead{
\colhead{Region} & \colhead{$k\tx$ [keV]}   &
\colhead{$A$}   & \colhead{red.\ $\chi^2$} & \colhead{D.O.F.}
}
\startdata
1 & $ 4.71^{+ 0.28}_{- 0.23}$ & $0.32^{+0.10}_{-0.10}$ &  1.06 &  259 \\[0.9mm]
2 & $ 4.54^{+ 0.27}_{- 0.26}$ & $0.33^{+0.10}_{-0.10}$ &  1.08 &  267 \\[0.9mm]
3 & $ 3.97^{+ 0.19}_{- 0.15}$ & $0.41^{+0.08}_{-0.07}$ &  0.96 &  439 \\[0.9mm]
4 & $ 4.70^{+ 0.27}_{- 0.24}$ & $0.22^{+0.09}_{-0.08}$ &  1.07 &  289 \\[0.9mm]
5 & $ 4.89^{+ 0.33}_{- 0.30}$ & $0.17^{+0.09}_{-0.09}$ &  0.94 &  316 \\[0.9mm]
6 & $ 3.94^{+ 0.27}_{- 0.23}$ & $0.27^{+0.10}_{-0.09}$ &  1.11 &  410 \\[0.9mm]
7 & $ 4.69^{+ 0.25}_{- 0.25}$ & $0.22^{+0.08}_{-0.08}$ &  0.83 &  318 \\[0.9mm]
8 & $ 4.11^{+ 0.27}_{- 0.27}$ & $0.24^{+0.10}_{-0.10}$ &  1.03 &  340 \\[0.9mm]
9 & $ 3.64^{+ 0.33}_{- 0.31}$ & $0.18^{+0.11}_{-0.11}$ &  0.93 &  413 \\[0.9mm]
\enddata
\end{deluxetable}
%
\begin{deluxetable}{ccccc}
\tabletypesize{\footnotesize}
\tablecaption{Fit results for the sub cluster regions \label{tbl:sreg}}
\tablewidth{0pt}
\tablehead{
\colhead{Region} & \colhead{$k\tx$ [keV]}   &
\colhead{$A$}   & \colhead{red.\ $\chi^2$} & \colhead{D.O.F.}
}
\startdata
1 & $ 5.24^{+ 1.79}_{- 1.12}$ & $0.70^{+1.34}_{-0.70}$ &  0.89 &   30 \\[0.9mm]
2 & $ 3.78^{+ 0.31}_{- 0.29}$ & $0.36^{+0.13}_{-0.13}$ &  0.92 &  217 \\[0.9mm]
3 & $ 3.18^{+ 0.24}_{- 0.22}$ & $0.59^{+0.14}_{-0.13}$ &  1.27 &  165 \\[0.9mm]
4 & $ 5.49^{+ 1.45}_{- 0.99}$ & $0.42^{+0.69}_{-0.42}$ &  1.46 &   31 \\[0.9mm]
5 & $ 3.09^{+ 0.41}_{- 0.37}$ & $0.31^{+0.24}_{-0.22}$ &  1.28 &   45 \\[0.9mm]
\enddata
\end{deluxetable}

\section{Discussion}\label{discu} 

We first consider some implications of these observations for the use
of clusters of galaxies as cosmological probes.
How much do the X-ray flux estimates based on resolved observations differ from 
simple estimates based on global measurements, such as those for clusters
only observed with the ROSAT All-Sky Survey (\ra)?
How much do mass estimates differ when only a broad beam overall gas temperature
estimate is available?

The overall flux from A1644, which extends beyond \xmm's
field-of-view, is $\fx(\eb)=4.03\esc$ \citep{rb01} based on \ra\ data.
(A1644 was not observed in pointed mode with the \ps.)
It is worth noting that A1644 has been treated as a single cluster in the \gcsa\
\citep{rb01} and REFLEX \citep{bsg01} catalogs, primarily due to the
limited exposure time ($<$ 300\,s) of the \ra\ at A1644's position and the
relatively small projected separation of the two cluster components.
The \xmm\ image indicates that the X-ray flux ($\fx$) ratio between the main and
sub clump is about 3:1. This 
means instead of one cluster with $\fx\approx 4\esc$ one actually has two
clusters with about $\fx\approx 3$ 
and $1\esc$ each.
This could affect luminosity and mass functions significantly if such
major mergers were common.
The effect of an incorrect treatment would be to artificially increase the
number of high mass clusters and decrease the number of low mass clusters.
Using the shape of the mass function to constrain the cosmic mass density,
$\om$, and the amplitude of mass fluctuations, expressed as $\sigma_8$,
individually would then result in artificially low values for $\om$ and high
values for $\sigma_8$.
With the ongoing analysis of the complete sample of \gcsa\ clusters with \cha\
and \xmm, we will be able to quantify the fraction of such clusters and their
influence on derived cosmological parameters,
as well as to study the difficulty of estimating survey selection functions in
the vicinity of bright clusters (not only due to double clusters but also line
of sight projections) in general.

A simple broad beam temperature estimate may be biased low compared to the
ambient gas temperature due to cool emission in the dense (high emissivity)
cores.
For instance, the temperature in a large annulus around the main clump
in A1644 (201--325$\arcsec$) is 4.40\,keV.
If we take this as the ambient temperature which is characteristic of
the cluster gravitational potential,
it is a factor of 1.15 higher than
a broad beam temperature estimate including both clumps (3.83\,keV).
Since $M\propto T^{1.5-2.0}$ \cite*[e.g.,][]{frb00}, this factor translates
into an underestimate of a factor of 1.23--1.32 in the cluster mass.

The surface brightness structure of both sub clumps is obviously messy.
However, the radial temperature profiles (Figs.~\ref{Tprofm} and \ref{Tprofs})
are smooth and appear rather similar to profiles of relaxed clusters
\citep*[e.g., compare to Fig.~1 in][]{asf01}.
This might suggest that A1644 is moderately relaxed, and that
the temperature structure may not be strongly affected by the
interaction of the two sub clumps.
However, a more detailed examination of the thermal structure shows that
the cluster is not at all relaxed.
As the HR map (Fig.~\ref{HR}) indicates and
Figs.~\ref{Tregm} and \ref{Tregs} confirm,
the temperature structure is also quite complex.
The region between the two clumps is significantly hotter than the region to the
west of the main clump
(the region which appears least disturbed in X-ray surface brightness).
This may indicate that the
gas in the region between the clumps has been heated up by adiabatic
compression or shocks. 
Also, the core-within-a-core structure of the main clump (Fig.~\ref{ixmm}) may
be caused by core oscillations induced by the passage of the sub clump
\citep[e.g., ][]{th03} or possibly by a remnant of a previous merger.

Furthermore, both the surface brightness and temperature structure around the
sub clump are peculiar.
Based on the X-ray image, the gas appears to be slightly compressed to the
north and clearly elongated to the south and west.
This suggests that this sub clump is moving
through the intracluster medium (ICM) of the main clump,
and losing some fraction of its gas in a trail behind the core of the sub
clump.
Given this dynamical picture what temperature structure might be expected?
One may expect heated gas in front of the sub clump (projected roughly to
the north) due to adiabatic compression or shocks.
One might also expect the trail to the south to consist of somewhat cooler,
more metal-rich gas stripped from the sub clump.
This appears to be consistent both with the HR map and with the extracted
spectra in the sub clump regions R1, R3, and R5.

Figs.~\ref{iopt} and \ref{irad2} show that the X-ray peak of the sub clump
and the optical center of the brightest galaxy coincide.
This suggests that the sub clump has not passed through the core of the
main clump; if it had, the very high ram pressure would have stripped the
sub clump core, which would lag behind the collisionless dark matter and
galaxies
\citep[e.g.,][]{mgd02}.
The angle of the cool, metal-rich gas trail also suggests an off-center
collision.
One concern with this picture is that the dense gas in the core of the
sub clump would then be interacting with lower density gas in the outer
parts of the main clump.
Would one expect the dense sub cluster to lose an observable amount of gas
during its travel through the ICM?

\begin{figure}[thbp]
\psfig{file=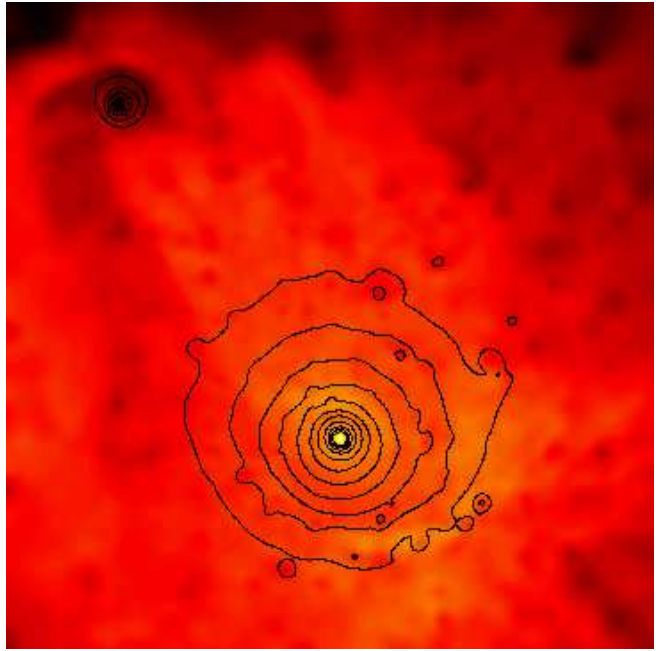,width=8.6cm,angle=0,clip=}
\caption{Mass weighted temperature map from a hydrodynamical simulation.
Temperature increases from black to red to yellow.
Also shown are X-ray surface brightness contours.
}\label{Tmap}
\end{figure}

We have compared A1644 to a hydrodynamical simulation of a small cluster
undergoing an off-axis merger with a larger cluster.
This simulation
was done completely independently; no effort was made to
adjust the initial conditions to match the observation of A1644.
The temperature map of this simulation at a time when the smaller cluster has
passed by the core of the main cluster for the first time is shown in
Fig.~\ref{Tmap}. 
Comparison to the HR map of A1644 (Fig.~\ref{HR}) shows an interesting
similarity.
There is hotter gas in front of the smaller cluster and cooler, ram-pressure
stripped gas behind.
This trail of stripped gas cools adiabatically as it expands.
There is a hotter region to the southwest, which is similar to region R4
seen in A1644.
The larger scale $\tx$ distribution around the main cluster in the
simulation is also similar to that observed in A1644.
There appears to be slightly cooler gas to the
southeast in the simulation due to stripped gas from the smaller cluster,
which agrees with the R8 region in the HR map and spectra.\footnote{
An animation of this simulation is available at http://A1644.dark-energy.net.
Note that
cool trails from sub clusters are commonly seen in simulations. Compare, e.g.,
to the simulations by \citet[ see also the movie at
http://www.astro.virginia.edu/coolflow/abs.php?regID=184]{mbl03}.}

The indication for an increased metal abundance in the cool trail of A1644's sub
clump makes
other scenarios like cooling of intracluster gas onto the moving sub
clump \citep[as seen in A1795 on much smaller scales,][]{fse01} more
difficult to reconcile with the data. Note that the radio source
in the center of the sub clump (Fig.~\ref{irad2}) may provide additional energy
for gas removal from central denser regions.

\section{Summary}\label{summa} 

\xmm\ imaging spectroscopy of the galaxy cluster A1644 has shown
two sub clusters undergoing an off-axis merger with remarkable detail.
The X-ray analysis is also supported by optical data on the cluster,
and the resulting dynamical picture agrees very well with an
independent hydrodynamical simulation of such an off-axis merger.
The findings imply that we see gas that has been
removed from the smaller sub cluster
possibly by ram pressure from the motion through intracluster gas of the
main clump, possibly augmented by the effect of the central radio source
of this sub clump.

\acknowledgements
The code to create the X-ray temperature map made
use of the Interactive Spectral Interpretation System (ISIS; \citealt{hd00}).
M. Ledlow made available the radio images.
This work was supported by NASA \xmm\ Grants NAG5-10075, NAG5-13088, and
NAG5-13737.
T. H. R. acknowledges support by the Celerity Foundation through a 
Post-Doctoral Fellowship.
The \xmm\ project is an ESA Science Mission with instruments and contributions
directly funded by ESA Member States and the USA (NASA).
The Digitized Sky Surveys were produced at the Space Telescope Science Institute
under U.S. Government grant NAG W-2166. The images of these surveys are
based on photographic data obtained using the Oschin Schmidt Telescope on
Palomar Mountain and the UK Schmidt Telescope. 
This publication makes use of data products from 2MASS, which is a joint project
of the University of Massachusetts and IPAC/Caltech, funded by NASA and NSF.


\end{document}